\documentclass[ reprint,amsmath,amssymb,
aps]{revtex4-2}
\usepackage[T1]{fontenc}
\usepackage{geometry}
\usepackage{xcolor}
\usepackage{babel}
\usepackage{units}
\usepackage{amsmath}
\usepackage{amssymb}
\usepackage{stmaryrd}
\usepackage{graphicx}
\usepackage{wasysym}
\usepackage[bookmarks=false,hidelinks=true]{hyperref}

\makeatletter

\IfFileExists{lmodern.sty}{\usepackage{lmodern}}{}
\makeatother

\begin{document}
\title{Chaotic scattering in ultracold atom-ion collisions}

\author{Meirav Pinkas}
% \email{Corresponding author: }
\thanks{meirav.pinkas@weizmann.ac.il}

\affiliation{Department of Physics of Complex Systems and AMOS, Weizmann Institute of Science,
Rehovot 7610001, Israel}

\author{Jonathan Wengrowicz}
\affiliation{Department of Physics of Complex Systems and AMOS, Weizmann Institute of Science,
Rehovot 7610001, Israel}

\author{Nitzan Akerman}
\affiliation{Department of Physics of Complex Systems and AMOS, Weizmann Institute of Science,
Rehovot 7610001, Israel}

\author{Roee Ozeri}
\affiliation{Department of Physics of Complex Systems and AMOS, Weizmann Institute of Science,
Rehovot 7610001, Israel}

\begin{abstract}
We report on signatures of classical chaos in ultracold collisions between a trapped ion and a free atom. Using numerical simulations, we show that the scattering dynamics can be highly sensitive to initial conditions for various mass ratios and trapping frequencies, indicating the onset of chaos. We quantify this chaotic dynamics by calculating its fractal dimension. We show that for a trapped $^{88}$Sr$^+$ ion and a free $^{87}$Rb atom chaotic dynamic appears under experimentally relevant conditions, and find its characteristic energy scale. The observation of classical chaos in atom-trapped-ion collisions suggests that signatures of quantum chaos might appear, for example, through a Wigner-Dyson distribution of collisional resonances.

\end{abstract}
\maketitle

Cooling particles to the ultracold regime allows for remarkable control over their interactions, for example, through collisional resonances \cite{Chin2010FeshbachResonances}. Nevertheless, collisions between ultracold particles can still be complex and chaotic. For instance, in collisions between magnetic lanthanide atoms, a dense spectrum of Feshbach resonances appears, with statistical signatures of quantum chaos \cite{Maier2015EmergenceChaotic, Frisch2014QuantumChaos}. As another example, the large number of rovibrational states that couple to the entrance channel in atom-diatom and diatom-diatom collisions lead to the formation of long-lived complexes with underlying chaotic dynamics \cite{Mayle2012StatisticalAspects, Mayle2013ScatteringUltracold, Mayle2012StatisticalAspects, Mayle2013ScatteringUltracold,Bause2023UltracoldSticky,Hu2019DirectObservation,Gregory2019StickyCollisions,Liu2020PhotoexcitationLonglived,Gersema2021ProbingPhotoinduceda,Bause2021CollisionsUltracold}. There, chaotic features were also shown by classical simulations \cite{Croft2014LonglivedComplexes, Croft2017LonglivedComplexes}. 

Hybrid systems of ultracold atoms and ions can take advantage of the high level of control over both species to detect interactions in the ultracold regime \cite{Tomza2019cold}. The atom-ion interaction is characterized by a long-range, $V(r)\sim -r^{-4}$, attractive potential. In addition, the ion is usually strongly confined by time-dependent electric fields in a Paul trap \cite{Leibfried2003}. The long-range potential together with the ion trapping fields strongly affect the collisional dynamics. First, the long-rang potential leads to a $s$-wave threshold temperature that is several orders of magnitude lower than is typically found in neutral atomic systems. Second, the time-dependent fields of the trap can increase the collision energy \cite{Cetina2012}. The presence of the trap breaks the translational invariance of the collision dynamics, which can lead to the formation of bound states \cite{Cetina2012,Hirzler2022TrapAssisted,Pinkas2023TrapassistedFormation}. These effects impede the observation of quantum effects in ultracold atom-ion collisions. However, these effects are reduced in experiments with heavy-ion light-atom pairs, in which quantum features were recently observed \cite{Feldker2020BufferGas,Weckesser2021}. Moreover, it was shown that, due to the long-range potential, quantum signatures also appear high above the s-wave limit \cite{Sikorsky2018PhaseLock,Cote2018}.

Here, using numerical simulations, we show that the classical ultracold scattering of a trapped ion and a free atom has chaotic properties. We first analyze a simplified one-dimensional problem, focusing on a collision in which the ion is confined by a static harmonic trap. We observe high sensitivity to the collision initial conditions, which keeps appearing at ever-decreasing scales with self-similar behavior. We quantify the degree of chaos by calculating the dynamics fractal dimension. We repeat this calculation for two and three dimensions and time-varying Paul-trap fields, and show that a high fractal dimension, and the onset of chaos, also appear under experimentally realistic conditions.

We consider a classical scattering problem of two particles: an ion of mass $m_i$ and an atom of mass $m_a$. The ion is trapped in either a static harmonic trap, or a radio-frequency (rf) trap. Both types of traps can be modeled by the potential $V_{trap}(x_{i})=\frac{m_i\Omega^{2}}{8}\left(a+2q\cos\left(\Omega t\right)\right)x_i^2$. For a radio-frequency trap, $\Omega$ is the rf drive frequency, and $a$ and $q$ are the dimensionless Mathieu parameters that depend on the static and oscillating electric potentials \cite{Leibfried2003}. The secular harmonic frequency is given by $\omega=\frac{\Omega}{2}\sqrt{a+q^2/2}$. To investigate a static harmonic trap, we set $a=4\omega^2/\Omega^2$ and $q=0$. 

The particles interact through an attractive charged-induced dipole interaction, $V(r)=-C_4/2r^4$, where $r$ is the particles' relative distance, and $C_4$ is the leading order dispersion coefficient. The interaction is repulsive on a length scale of a few Bohr radii due to chemical forces \cite{Tomza2019cold}. We model this repulsion by the repulsive potential $V_{rep}(r)=C_6/r^6$.

We normalize the equations of motion by the time scale $t_{c}=\Omega^{-1}$ and length scale $x_{c}=(2C_{4}/m_{i}\Omega^{2})^{1/6}$ \footnote{See Supplementary Material for trajectory calculation, calculation of the uncertainty exponent and maximum likelihood estimation of the power-law.}, yielding 

\begin{equation}
\begin{split}
    \ddot{\chi}_{i}	& = -\nabla \tilde{V}(\left|\chi_{i}-\chi_{a}\right|)-\frac{1}{4}\left(a+2q\cos\left(\tau\right)\right)\chi_{i} \\
    \ddot{\chi}_{a}	& = \xi\nabla \tilde{V}(\left|\chi_{i}-\chi_{a}\right|),
\end{split}
\label{eq:nondim_eom}
\end{equation}

where $\chi_{i}$($\chi_{a}$) is the position of the ion (atom), $\tilde{V}$ is the dimensionless interaction potential, and $\xi=m_{i}/m_{a}$ is the ion-atom mass ratio.

\begin{figure}
    \centering
    \includegraphics[width=0.48\textwidth]{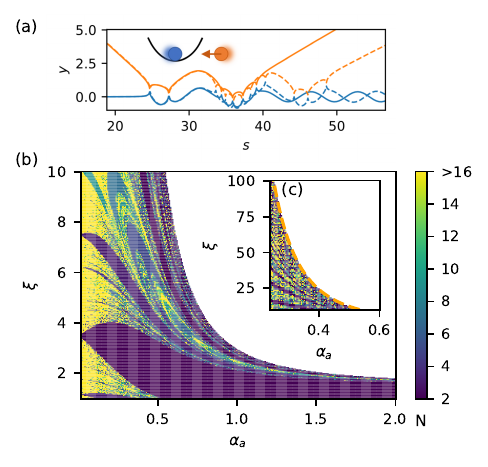}
    \caption{{An illustration of the 1D collision set-up, trajectories, and number of collisions as a function of collision parameters.} (a) A trapped ion and a free atom bound state. The solid blue (orange) line is the ion (atom) trajectory, calculated by solving Eq.~\ref{eq:scaled_nondim_harmonic_eom} with $\xi = 1.011, \alpha_a\equiv\dot{y}_a=0.4988$. The dashed line trajectory is calculated after changing $\alpha_a$  by less than 0.5\%. (Inset) A trapped ion (blue) in a harmonic trap, and a free atom (red) with an initial velocity $\alpha_a$. (b) The number of short-range collisions for the 1D problem (Eq.~\ref{eq:scaled_nondim_harmonic_eom}) as a function of the mass ratio, $\xi$, and scaled dimensionless initial atom's velocity, $\alpha_a$. The ion is initialized with zero energy. (c) Same as (b) but for higher values of $\xi$.}
    \label{fig:1D_scheme_and_map}
\end{figure}

In the case of a one-dimensional (1D) problem with the ion in a static harmonic trap ($q=0$), Eq.~\ref{eq:nondim_eom} can be scaled by the change of variables $s=\frac{\sqrt{a}}{2}\tau$ and $y=\sqrt[6]{\frac{a}{4}}\chi$, giving

\begin{equation}
    \ddot{y}_{i}	 = -\frac{y_{i}-y_{a}}{\left|y_{i}-y_{a}\right|^{6}}-y_{i},\quad \quad
    \ddot{y}_{a}	 = \xi\frac{y_{i}-y_{a}}{\left|y_{i}-y_{a}\right|^{6}}.
\label{eq:scaled_nondim_harmonic_eom}
\end{equation}
In the solution of these equations, the mass ratio $\xi$ and the particles' initial conditions set the trajectories.
A schematic drawing of the problem in 1D, and two trajectory examples are shown in Fig.~\ref{fig:1D_scheme_and_map}(a).

In free space, two particles cannot be bound in an elastic collision due to energy and momentum conservation. However, the trapping potential can couple the center-of-mass and relative motion, and hence the particles interact at short range several times before separating \cite{Pinkas2023TrapassistedFormation, Hirzler2022TrapAssisted}. The example trajectories in Fig.~\ref{fig:1D_scheme_and_map}(a) contain a few short-range interactions. When perturbing the initial condition even by a small amount, the number of short-range collisions changes significantly, as shown by the dashed lines in Fig.~\ref{fig:1D_scheme_and_map}(a).

To study this sensitivity to initial conditions, we calculate the number of close contact collisions as a function of the mass ratio $\xi$, and the scaled initial atom velocity, $\alpha_a$, shown in Fig.~\ref{fig:1D_scheme_and_map}(b-c). In this calculation, the ion has zero energy, and the atom starts far away from the ion. Since this is a 1D problem, the particles always have at least one close contact. We define a bound state whenever the particles have multiple short-range collisions. For each mass ratio, the particles always create a bound state below a critical atomic velocity, $\alpha_\textrm{a}^\textrm{c}(\xi)$.

In the large mass-ratio limit, $\alpha_\textrm{a}^\textrm{c}(\xi)$ can be found empirically from Fig.~\ref{fig:1D_scheme_and_map}(c) by fitting the borderline between single and multiple collisions to the function $\alpha_c (\xi) = c\xi^{-n}$  yielding  $n=0.3277(5)$ and $c=1.135(5)$. Approximating this relation by $\alpha_c (\xi) \sim \xi^{-1/3}$, the critical kinetic energy for creating a bound state is given by
\begin{equation}
    E_a^c \sim \frac{1}{2}\left(\frac{2C_{4}\omega^4m_a^5}{m_{i}^3}\right)^{1/3}.
    \label{eq:E_scale}
\end{equation}
We note that this energy scale resembles the radio-frequency heating energy scale in ref. \cite{Cetina2012}. However, here, the trap is static, without time-dependent fields. Therefore, bound state formation appears at a similar energy scale, even with an energy-conserving trap.

\begin{figure}
    \centering
    \includegraphics[width=0.49\textwidth]{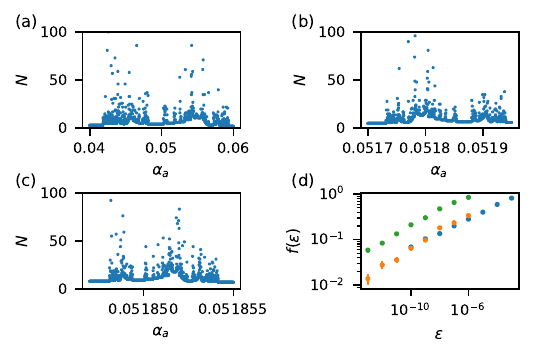}
    \caption{Sensitivity to initial condition and uncertainty exponent for a $^{88}\mathrm{Sr}^+-{}^{87}\mathrm{Rb}$ collision in 1D ($\xi$=1.0114). (a-c) Number of short-range collisions as a function of the dimensionless initial velocity of the atom, $\alpha_a$. (d) Uncertainty functions for (a-c) in blue, orange, and green, as a function of the perturbation $\epsilon$, giving uncertainty exponents of $\alpha=0.153(2),\ 0.193(2)$ and $0.168(1)$, respectively.}
    \label{fig:1D_self_sim}
\end{figure}

When a bound state is formed, for some parameter regions the number of short-range collisions is constant, whereas in others, it changes abruptly as a function of the collision parameters. To examine whether this dynamics is regular or chaotic, we focus on the case of  $^{88}$Sr$^+$-$^{87}$Rb pair ($\xi$=1.0114).
Fig.~\ref{fig:1D_self_sim}(a) shows the number of short-range collisions as a function of the initial velocity of the atom, which exhibits high sensitivity to the initial conditions in some regions. This large sensitivity keeps appearing for higher resolution of the initial conditions, as shown in Fig. \ref{fig:1D_self_sim}(b-c). Self-similarity is a feature of fractals, which appear frequently in chaotic systems, for example, as fractal basin boundaries \cite{Mcdonald1985FractalBasin}.

Chaotic systems are usually characterized by the exponential separation of nearby trajectories over a long evolution time, quantified by the Lyapunov exponent. However, chaos can also appear in transient dynamics \cite{Lai2011TransientChaos}. For example, in scattering problems of a particle in a hard-spheres or hills potentials, the deflection angle and the delay time of the particle can have a large sensitivity on the initial conditions, and demonstrate a fractal nature \cite{Gaspard1989ScatteringClassically, Bleher1990BifurcationChaotic}.

Singularities in the scattering function can be quantified by calculating their fractal dimension. The fractal dimension, $d$, quantifies how measured properties change when the scale over which they are measured is changed \cite{Mandelbrot1967HowLong, Lai2011TransientChaos}. An equivalent method is by determining the uncertainty exponent $\alpha$ \cite{Mcdonald1985FractalBasin}, which quantifies the stability of a trajectory under a perturbation $\epsilon$ of the initial condition.

Here, we define a trajectory to be stable if the number of short-range collisions is unchanged under a small perturbation. To this end, we find the uncertainty exponent for the data in Fig.~\ref{fig:1D_self_sim}(a-c), by choosing a random initial atom velocity, $\alpha_a$, and calculating the trajectories for two close velocities: $\alpha_a$ and $\alpha_a+\epsilon$. By repeating this process for different values of $\alpha_a$ we can calculate the fraction of unstable initial conditions, $f(\epsilon)$; i.e. the fraction of cases in which the number of short-range collisions is different. Non-chaotic basin boundaries obey $f(\epsilon)\sim \epsilon$ whereas chaotic boundaries obey $f(\epsilon) \sim \epsilon^\alpha$, where $\alpha$ is known as the uncertainty exponent \cite{Mcdonald1985FractalBasin}. For a perturbation in a single dimension, the fractal dimension $d$ is given by $d=1-\alpha$.

We first focus on the data in Fig.~\ref{fig:1D_self_sim}(a-c). For the 1D system, the singularities are points, i.e. $d=0$. However, if fractal, the dimension can be higher, i.e., $0<d<1$. The uncertainty function $f(\epsilon)$ for the region in Fig.~\ref{fig:1D_self_sim}(a) is shown in blue in Fig.~\ref{fig:1D_self_sim}(d), calculated using $10^3$ repetitions. A Maximum Likelihood Estimation (MLE) for the uncertainty exponent yields $\alpha=$0.153(2) \cite{Note1}. Similar values are also observed for the regions in Fig.~\ref{fig:1D_self_sim}(b-c), shown in Fig.~\ref{fig:1D_self_sim}(d) in orange and green, respectively. This value means that reducing the perturbation by a factor of 100, improves the predictability only by a factor of two. Since the perturbation is 1D, the fractal dimension for the region in Fig.~\ref{fig:1D_self_sim}(a) is $d=0.847$. This value significantly differs from the expected non-chaotic value of $d=0$, indicating a strong chaotic behavior. 

\begin{figure}
    \centering
    \includegraphics[width=0.499\textwidth]{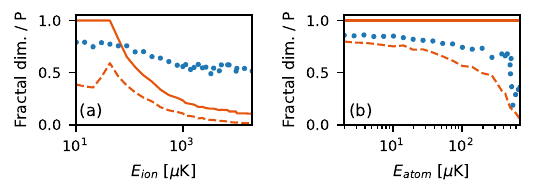}
    \caption{The fractal dimension (blue) and a bound state probability (red) as a function of (a) initial ion's energy ($T_{atom} = 100\ \mu K\cdot k_\mathrm{B}$) and (b) atom's energy ($T_{ion} = 50\ \mu K\cdot k_\mathrm{B}$). 1$\sigma$ confidence bounds are smaller than marker size. The orange solid (dashed) line is the probability of more than two (three) short-range collisions.}
    \label{fig:fractal_dimenergy}
\end{figure}

A transition between chaotic and regular scattering can occur at specific energies \cite{Bleher1989RoutesChaotic, Bleher1990BifurcationChaotic}. In our case, we study two different cases for initial kinetic energy. One originates in the ion's harmonic motion, and the other is due to the atom's initial velocity. Fig.~\ref{fig:fractal_dimenergy} shows the fractal dimension as a function of each of these initial kinetic energies. Introducing energy to the ion sets the initial amplitude of the harmonic oscillator, but not the initial phase. The uncertainty function $f(\epsilon)$ here is calculated by perturbing the initial phase by $\epsilon$, while maintaining the same initial energy. 

Fig.~\ref{fig:fractal_dimenergy} shows that chaotic scattering, similar to molecular formation, is a hallmark of low energies. However, the behavior differs when the initial energy is in atom or ion motion. In the case of an initially moving $^{87}$Rb atom, the typical onset of chaos, marked by a rise in the fractal dimension, is around $\sim$ 0.5 mK. However, the change to the fractal dimension around 0.5 mK in the case where the $^{88}\mathrm{Sr}^+$ ion is initially in motion is only very slight. This is because, in the latter case, the energy is in harmonic motion and, depending on the phase of motion at the moment of collision, the momentary energy during the collision can still be low, leading to chaos and a bound state formation. 

In addition, we can see that the fractal dimension is reduced and non-chaotic behavior is restored on a similar energy scale to that of bound state formation, shown by the solid and dashed orange lines in Fig.~\ref{fig:fractal_dimenergy}(a). Chaotic behavior is seen to be closely related to bound state formation.

\begin{figure}
    \centering
    \includegraphics[width=0.49\textwidth]{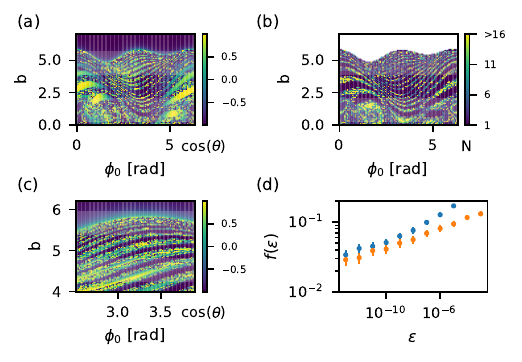}
    \caption{Sensitivity of collision trajectories to initial conditions in 2D. (a) Cosine of the scattering angle, $\theta$, of the atom and (b) the number of short-range collisions for a two-dimensional problem as a function of dimensionless impact parameter $b$, and initial phase of the ion $\phi_0$. The ion is harmonically trapped ($\omega/2\pi$ = 1 MHz) with initial kinetic energy of 50 $\mu K\cdot k_\mathrm{B}$, with initial velocity perpendicular to the initial velocity of the atom. The atom's initial kinetic energy is 100 $\mu K\cdot k_\mathrm{B}$. (c) The same as (a) but zooming on a smaller region. (d) Fraction of unstable trajectories as a function of the perturbation $\varepsilon$ for the regions shown in (a) and (c), in blue and orange respectively, giving uncertainty exponents of 0.092(3) and 0.068(3).}
    \label{fig:2D_self_sim}
\end{figure}

We now move to discuss atom-ion collisions in two dimensions (2D). While in the 1D atom-ion collision problem, the atom always collides with the ion, in the 2D problem, the occurrence of a short-range collision depends on the impact parameter, $b$. We solve the $^{87}$Rb atom $^{88}$Sr$^+$ ion trajectories in 2D where the ion is confined by a static harmonic trap with non-zero initial energy. Here we study the cosine of the scattering angle of the atom following the collision, $\cos(\theta)$, shown in Fig.~\ref{fig:2D_self_sim}(a), and the number of short-range collisions, shown in Fig.~\ref{fig:2D_self_sim}(b).

For impact parameters below the critical value, the scattering angle presents sensitive and non-sensitive regions. Regions that are smooth in terms of $\cos(\theta)$ in Fig.~\ref{fig:2D_self_sim}(a) are also constant in terms of the number of short-range collisions in Fig.~\ref{fig:2D_self_sim}(b). Irregular regions agree in both cases as well. Self-similarity is also evident in this case, revealing stripes of regular and irregular regions, shown in Fig.~\ref{fig:2D_self_sim}(c). We calculate the uncertainty exponent similarly to the 1D case, i.e. by defining the stability with respect to the number of short-range collisions. For the 2D region in Fig.~\ref{fig:2D_self_sim}(c), f($\epsilon$) is shown in Fig.~\ref{fig:2D_self_sim}(d), giving a slightly lower value as compared with the 1D case, $\alpha$ = 0.068 ($d=$0.93).

In most hybrid atom-ion experiments, the ion is trapped using oscillating radio frequency electric fields in 3D. These fields change the collision dynamics, for example, by heating the ion \cite{Cetina2012}. To check whether chaotic signatures remain when the ion is trapped in a 3D Paul trap, we solve the equations of motion for $^{88}\mathrm{Sr}^{+}-{}^{87}\mathrm{Rb}$ experimental conditions \cite{Meir2017ExperimentalApparatus,Note1}.

We calculate the fractal dimension for the 3D problem by perturbing the initial position of the atom. We find an uncertainty exponent $\alpha=0.131(2)$ and hence a fractal dimension of $d=0.87$, similar to the 1D harmonic case. Therefore, also in the 3D Paul trap case, dynamics is still highly chaotic.

\begin{figure}[t]
    \centering
    \includegraphics[width=0.499\textwidth]{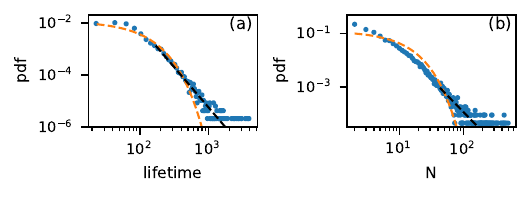}
    \caption{Distributions of lifetime and number of collisions for $^{88}$Sr$^+$-$^{87}$Rb experimental conditions in a Paul trap in 3D. (a) The dimensionless lifetime probability density function. The dashed orange line is a fit to an exponential distribution, yielding $\langle\tau \rangle$=80. Black dashed line is a fit to a power-law distribution with $\tau_{min}$=230(40) and $\alpha$=3.19(8). (b) Probability density function of the number of short-range collisions. The dashed orange line is a geometric distribution with $p$=0.11. Black dashed line is a power-law distribution with $N_{min}$=43(15) and $\alpha$=2.5(1). The parameters for both graphs and fitting procedures are described in \cite{Note1}.}
    \label{fig:3D_ensemble}
\end{figure}

We further calculate the distributions of bound state lifetimes and the number of short-range collisions over an ensemble of trajectories.
The lifetime distribution is shown in Fig.~\ref{fig:3D_ensemble}(a) in dimensionless units (corresponds to Eq.~\ref{eq:nondim_eom}). The maximal likelihood fit to an exponential distribution, $p(\tau) = \langle \tau \rangle^{-1} \exp{(-\tau/\langle \tau \rangle)}$, is shown by the orange solid line. The numerical distribution has a higher probability for long-lived events than predicted by an exponential distribution. The exponential model can be compared to a power-law model above a certain cutoff lifetime $\tau_{min}$ \cite{Note1}. The fit yields a power-law exponent of $\alpha=3.19(8)$. The distribution of the number of collisions yields a geometric distribution, $\mathrm{Geo}(k;p)=(1-p)^{k-2} p$ (defined for $k\ge2$), with a power-law exponent of $\alpha=2.5(1)$. In both cases, the likelihood of the power-law model is orders of magnitudes larger than the exponential or geometrical model above the critical value.

In chaotic scattering problems, the lifetime distribution usually obeys exponential decay \cite{Lai2011TransientChaos}. However, the lifetime distribution can show algebraic decay, if stable trajectories exist \cite{Yalcinkaya1995ChaoticScattering}. With stable trajectories, particles with chaotic trajectories can spend a long time in the vicinity of the stable region, which leads to algebraic decay of lifetime. For the trapped-ion and free atom case, the trap might stabilize trajectories, leading to long "sticking" time and hence algebraic decay of the lifetime. Recently, nonexponential decay was also suggested for molecular complexes \cite{Croft2023AnomalousLifetimes}.

To summarize, the scattering of an atom by a trapped ion leads to chaotic dynamics at low energies. For Sr$^{+}$-Rb collisions, this behavior appears both for an ion in a static harmonic trap or in a Paul trap with time-dependent fields. The existence of both regular and irregular regions result in a lifetime and a number of collisions distributions that exhibit a power-law tail, with a relatively high probability for long-lived molecular states.

The long-lived events might give an experimental signature of the chaos in the system. For example, detecting long-lived states can be done by a change in the rate of photo-excitation, photo-association processes, and three-body recombination \cite{Hirzler2022TrapAssisted}, relative to non-chaotic regions. Using harmonic potentials for trapping the ion, such as optical trapping \cite{Lambrecht2017,Schmidt2020OpticalTraps}, even longer lifetimes can be achieved \cite{Cui2023ColdHybrid}.

Chaos commonly appears in three-body gravitational problems, as was first examined by Poincare \cite{Poincare1892MethodesNouvelles}. An example of three-body gravitational scattering problem is the motion of an incoming star toward a bound pair of stars \cite{Hut1983TopologyThreebody,Boyd1992InitialvalueSpace, Boyd1993ChaoticScattering}. This motion exhibits a stripes structure of regular and chaotic scattering, similar to the one shown in Fig.~\ref{fig:2D_self_sim}(b). This structure arises due to the oscillatory motion in the problem (rotating bound pair or a trapped ion). Although the atom-trapped-ion problem is a two-body problem, the trap can be considered a massive third body to which the ion is bound.

Although the treatment in this work is classical, quantum chaos can arise in systems that exhibit chaotic behavior in the classical limit \cite{Bohigas1984CharacterizationChaotic}. Quantum chaos can appear as dense resonances with nearest-neighbor level spacing obeying Wigner-Dyson distribution, as observed in Feshbach resonances spectrum of magnetic lanthanides \cite{Maier2015EmergenceChaotic, Frisch2014QuantumChaos} and nuclear spectra \cite{Weidenmuller2009RandomMatrices}. Therefore, the spectra of atom-trapped-ion collisional resonances in the quantum regime might exhibit similar properties. In addition, in magnetic lanthanides, for example, the chaotic spectrum is intrinsic to the system. However, in the atom-trapped-ion system, it depends on the trapping potential. Therefore, the trapping potential can be tuned experimentally to chaotic or regular scattering regimes. In recent years, quantum atom-ion scattering was observed in the high-mass ratio regime \cite{Feldker2020BufferGas,Weckesser2021}. For these systems, the critical atomic energy estimation (Eq. \ref{eq:E_scale}) is on the order of $\sim$100 nK, which is below the experimental atomic temperature. Tuning the trapping frequency might also lead to observing chaos in these systems. However, substantial chaotic scattering could hinder the observation of shape resonances, which are expected for free-free collisions \cite{Sikorsky2018PhaseLock,Cote2018}.

The emergence of chaos here is closely related to trap-induced bound states formation \cite{Hirzler2022TrapAssisted,Pinkas2023TrapassistedFormation}. A similar mechanism - inelastic confinement-induced resonances - appears in ultracold atoms in unharmonic potentials \cite{Sala2013CoherentMolecule, Capecchi2022ObservationConfinementinduced,Kestner2010} and heteronuclear collisions \cite{Ruttley2023FormationUltracold,Melezhik2009QuantumDynamics}. These resonances are created by coupling center-of-mass and relative motion, leading to avoided crossings. On the other hand, quantum chaos statistics appear due to level repulsion \cite{Weidenmuller2009RandomMatrices}. This suggests that signatures of chaos might also appear when inelastic confinement-induced resonances are observed.

\hfill 

\begin{acknowledgments}
We thank Boaz Katz, Doron Kushnir, John Bohn, Or Katz, Oren Raz, Eleanor Trimby, and Rene Gerritsma for fruitful discussions. This work was supported by the Israeli Science Foundation Grant number 1385/19.
\end{acknowledgments}

\bibliographystyle{apsrev4-1}

\bibliography{refs}

\clearpage
\onecolumngrid
\part*{Supplementary Material}
\setcounter{equation}{0}
\setcounter{figure}{0}
\setcounter{table}{0}
\setcounter{section}{0}

\section{Trajectory calculation}
The equations of motion (EOM) of a trapped ion and free atom are

\begin{equation}
\begin{split}
    \ddot{x}_{i}	& =-\frac{1}{m_{i}}\left[ \frac{2C_{4}\left(x_{i}-x_{a}\right)}{\left|x_{i}-x_{a}\right|^{6}}-\frac{6C_6\left(x_{i}-x_{a}\right)}{\left|x_{i}-x_{a}\right|^{8}}\right]-\frac{\Omega^{2}}{4}\left(a+2q\cos\left(\Omega t\right)\right)x_{i} \\
\ddot{x}_{a}	& =\frac{1}{m_{a}}\left[ \frac{2C_{4}\left(x_{i}-x_{a}\right)}{\left|x_{i}-x_{a}\right|^{6}}-\frac{6C_6\left(x_{i}-x_{a}\right)}{\left|x_{i}-x_{a}\right|^{8}}\right],
\end{split}
\end{equation}
where $x_i$ ($x_a$) is the ion (atom's) position, $m_i$ ($m_a$) is the ion's (atom) mass,  $\Omega$ is the rf drive frequency, $a$ and $q$ are the trapping parameters \cite{Leibfried2003}, $C_4$ is the leading order dispersion coefficient for the attractive atom-ion interaction, and $C_6$ represents the short-range repulsion coefficient. For simplicity, the equations here are written for the 1D case, but this can be generalized to higher dimensions by taking $x_i$, $x_a$, $a$, and $q$ as vectors.

For the dimensionalization process, we choose the characteristic timescale, $t_{c}=\Omega^{-1}$ and length scale, $x_{c}=\sqrt[6]{2C_{4}/(m_{i}\Omega^{2})}$. The dimensionless time and position are defined as, $\tau=t/t_{c}$ and $\chi=x_{i}/x_{c}$, respectively. The dimensionless EOM are

\begin{equation}
\begin{split}
    \ddot{\chi}_{i}	& = -\frac{\left(\chi_{i}-\chi_{a}\right)}{\left|\chi_{i}-\chi_{a}\right|^{6}} +\frac{3}{2}\beta^{2}\frac{\left(\chi_{i}-\chi_{a}\right)}{\left|\chi_{i}-\chi_{a}\right|^{8}} -\frac{1}{4}\left(a+2q\cos\left(\tau\right)\right)\chi_{i} \\
    \ddot{\chi}_{a}	& = \xi\left[\frac{\left(\chi_{i}-\chi_{a}\right)}{\left|\chi_{i}-\chi_{a}\right|^{6}}-\frac{3}{2}\beta^{2}\frac{\left(\chi_{i}-\chi_{a}\right)}{\left|\chi_{i}-\chi_{a}\right|^{8}}\right]
\end{split}
\label{eq:full_nondim_eom}
\end{equation}
where $\beta = x_{c}^{-1}\sqrt{2C_6/C_{4}}$. Using this notation, the dimensionless interaction potential is $\tilde{V}(\rho)=-\frac{1}{4\rho^4}+\frac{\beta^2}{4\rho^6}$. $\beta$ can be expressed using the classical turning point $r_{turn}$ as $\beta = r_{turn}/x_{c}$. We choose $\beta = 0.76$, which corresponds to a turning point of 10 nm for the Rb $C_4$ coefficient. For $^{88}$Sr-$^{87}$Rb, the parameters are given in Table \ref{tab:params}, and the typical length- and time-scale are $x_c$ = 13 nm and $t_c$ = 6 ns, respectively.

For the 1D trajectory and a harmonic trap for the ion ($q$ = 0), the equations can be simplified by an additional scaling. Choosing $s=\frac{\sqrt{a}}{2}\tau$ and $y=\sqrt[6]{\frac{a}{4}}\chi$, gives

\begin{equation}
\begin{split}
    \ddot{y}_{i}	& =-\frac{\left(y_{i}-y_{a}\right)}{\left|y_{i}-y_{a}\right|^{6}}+\frac{3}{2}\tilde{\beta}^{2}\frac{\left(y_{i}-y_{a}\right)}{\left|y_{i}-y_{a}\right|^{8}}-y_{i} \\
\ddot{y}_{a}	& = \xi\left[\frac{\left(y_{i}-y_{a}\right)}{\left|y_{i}-y_{a}\right|^{6}}-\frac{3}{2}\tilde{\beta}^{2}\frac{\left(y_{i}-y_{a}\right)}{\left|y_{i}-y_{a}\right|^{8}}\right], \\
\end{split}
\end{equation}
where $\tilde{\beta} =\sqrt[6]{\frac{a}{4}} \beta$. $\tilde{\beta}$ corresponds to the classical turning point in the scaled units.

The differential equations are solved in Julia language using the 12th order explicit adaptive Runge-Kutta-Nyström algorithm (DPRKN12) with a relative tolerance of 10$^{-15}$. The atom starts far away from the ion,  $\chi_a(0)$ > 20. The integration stops when the atom is farther than its initial condition, or when $\tau$>5000.

For the 3D simulation, the initial conditions are random and sampled as described in Ref.~\cite{Pinkas2020}, with the corresponding dimensionless values. The Sr$^+$ ion is trapped in a linear Paul trap, 
where the trapping parameters are $a=(-1.3,-1.3,2.6)\times10^{-3}$, $q=(0.13,-0.13,0)$, and driving frequency $\Omega/(2\pi)=26.51$ MHz. This corresponds to radial trapping frequencies of $\omega_{rad}/(2\pi) = 1.12$ MHz and axial frequency of $\omega_{ax}/(2\pi) = 0.676$ MHz. The initial energy of the ion is zero, and therefore $\chi_i^j(0), \dot{\chi}_i^j(0)=0$, where $j$ denote the $j$-th coordinate.

The initial position of the atom is uniformaly distributed on a plane at a distance 40 from the equilibrium position of the ion, i.e.  $\chi_a^x,\chi_a^z \in \text{U}(-40,40)$ and $\chi_a^y=40$. The initial velocity of the atom has a constant contribution $\alpha_a=\sqrt{\frac{2k_B E_{lat}}{m_a}}\frac{t_c}{x_c}$ in the lattice direction, and a thermal contribution $\alpha_a^T=\sqrt{\frac{k_B T_a}{m_a}}\frac{t_c}{x_c}$, where $E_{lat}$ is the kinetic energy of atoms in the lattice and $T_a$ is the temperature of the atoms. In the direction perpendicular to the plane (y), the initial velocity is given by $\dot{\chi}_a^y(0)\in \alpha_a + \text{Rayleigh}(\alpha_a^T)$, where Rayleigh(...) is the Rayleigh distribution. On the parallel directions to the plane, the initial velocities are given by the Normal distribution $\dot{\chi}_a^x,\dot{\chi}_a^z\in N(0, \alpha_a^T)$. For this simulation, $\alpha_a$ = 6.31$\cdot10^{-2}$, and $\alpha_a^T$=9.98$\cdot 10^{-3}$ which corresponds to a lattice energy of 100 $\mu$K, and a temperature of 5 $\mu$K, respectively. The total number of repetitions is $10^5$.

\section{Calculation of the uncertainty exponent}
To calculate the uncertainty exponent, many trajectories are calculated with random initial conditions. For the 1D and 2D cases, the initial conditions are taken to be uniformaly distributed in a selected region, and all other parameters are constant. For the 3D experimental case, initial conditions are as described in the previous section. Then, for each trajectory, the initial condition is perturbed by $\epsilon$. For the 1D case, the initial atom's velocity is perturbed (as in Fig.~\ref{fig:1D_self_sim}(d)), or the initial phase of the ion (as in Fig.~\ref{fig:fractal_dimenergy}). For the 2D case (as in Fig~\ref{fig:2D_self_sim}(c)), the initial ion's phase is perturbed. The number of collisions is compared for each perturbed trajectory with the unperturbed one. The trajectory is considered unstable under perturbation $\epsilon$ if the number of collisions differs. For each $\epsilon$, the fraction of unstable trajectories, $f(\epsilon)$, is calculated.

The fraction $f(\epsilon)$ is expected to have a power-law dependence for fractal boundaries \cite{Mcdonald1985FractalBasin}. We find the corresponding power-law exponent by maximum likelihood estimation. We assume that for each $\epsilon$, $f(\epsilon)$ has a binomial distribution with probability $p(\epsilon)=C\epsilon^{-\alpha}$ with $m$ trails and $n=mf(\epsilon)$ successes. The logarithm of the likelihood function is
\begin{equation}
    \mathcal{L}(\alpha,C) = \sum_{\epsilon_i}\left[ n \log(p(\epsilon_i))+(m-n)\log(1-p(\epsilon_i)) + \log\binom{m}{n}\right].
\end{equation}

We find the parameters $(\hat{\alpha}, \hat{C})$ that maximized the likelihood function. The last term is independent of the parameters and is neglected in the optimization process. The confidence interval of $\alpha$ is calculated from the log-likelihood function by finding $\alpha$ such that
\begin{equation}
    \alpha_{ci} = \{ \alpha |\mathcal{L}(\hat{\alpha},\hat{C}) - \mathcal{L}(\alpha,\hat{C}) = \text{invcdf}_{\chi^2}(q,\nu) \}
\end{equation}
where $\text{invcdf}_{\chi^2}(q,\nu)$ is the chi-square inverse cumulative distribution function for $q$ degrees of freedom at probability $\nu$. For two degrees of freedom MLE and $1\sigma$ confidence bound, $q=2$ and $\nu$=0.68.

For the dependence of the fractal dimension on the atom's energy (Fig.~\ref{fig:fractal_dimenergy}(b)) at large energies above $\sim$ 500 $\mu$K, the $f(\epsilon)$ dependence no longer obeys a power law for large $\epsilon$ and only points with $\epsilon<10^{-1}$ are taken.

\section{Maximum likelihood estimation (MLE) of the power-law}
We follow the methods described in Ref.~\cite{Clauset2009PowerLawDistributions} to find the power-law exponent and corresponding cutoff $x_{min}$. 

\subsection{Continuous distribution}
We assume the lifetime distribution has as a power-law $\alpha$ above $x_{min}$, 
\begin{equation}
    p(x) = (\alpha - 1) x_{min}^{\alpha - 1}x^{-\alpha}.
\end{equation}
This probability is normalized such that 
\begin{equation}
    \int_{x_{min}}^\infty p(x)= 1
\end{equation}

The estimator for the power-law $\hat{\alpha}$ is given by \cite{Clauset2009PowerLawDistributions}
\begin{equation}
    \hat{\alpha} = 1+n\left[\sum_{i=1}^n \ln \frac{x_i}{x_{min}} \right]^{-1},
\end{equation}
where $x_i$ are all observed lifetime that satisfy $x_i>x_{min}$. To find the cutoff time $x_{min}$, we minimize the Kolmogorov–Smirnov statistic
\begin{equation}
    D = \max_{x\ge x_{min}} \left| S(x) - P(x) \right|
\end{equation}
where $S(x)$ is the empirical CDF of that observed data with $x\ge x_{min}$ and $P(x)$ is the CDF of the fitted power-law model for $x\ge x_{min}$.
We find the standard deviation of $\alpha$ and $\tau_{min}$ by bootstrapping the lifetime distribution $10^3$ times and repeating the process above.

\subsection{Discrete distribution}
The number of collisions is a discrete distribution. A discrete power law distribution above $x_{min}$ can be  described by the probability density distribution
\begin{equation}
    p(x) = \frac{x^{-\alpha}}{\zeta(\alpha, x_{min})}
\end{equation}

where
\begin{equation}
    \zeta(\alpha, x_{min}) = \sum_{n=0}^\infty (n+x_{min})^{-\alpha}.
\end{equation}

We numerically find the estimator $\hat{\alpha}$, by maximizing the logarithm of the likelihood function for a given $x_{min}$
\begin{equation}
    \mathcal{L}(\alpha) = n\log \zeta(\alpha, x_{min}) - \alpha \sum_{i=1}^n \log x_i
\end{equation}
where $n$ is the sample size. The cutoff and the standard deviation of $\alpha$ are found in the same process as for the continuous distribution.

\begin{table}[]
    \centering
\begin{tabular}{||c | c ||} 
 \hline
 Parameter & Value  \\ [0.5ex] 
 \hline\hline
 $C_4$ [J$\cdot$m$^4$ ]& 1.05e-56    \\ 
 \hline
 $m_{i}$ [amu] & 87.905612257    \\
 \hline
 $m_{a}$ [amu] & 86.909180520   \\
 \hline
 $\Omega$ [2$\pi\cdot$MHz] & $2\pi\cdot$26.51  \\
 \hline
\end{tabular}
    \caption{Physical parameters for $^{88}$Sr$^{+}$-$^{87}$Rb simulation.}
    \label{tab:params}
\end{table}

\begin{center}

\end{center}

\clearpage
\setcounter{equation}{0}
\setcounter{figure}{0}
\setcounter{table}{0}

\def\equationautorefname~#1\null{Equation (#1)\null}
\newcommand{\diff}{\mathrm{d}}

\renewcommand{\theequation}{S\arabic{equation}}
\renewcommand{\thefigure}{S\arabic{figure}}
\renewcommand{\thetable}{S-\Roman{table}}

\onecolumngrid 
\setcounter{page}{1}
\end{document}